\documentclass[reprint,pra,english,superscriptaddress,floatfix,longbibliography,amsmath,amssymb,aps]{revtex4-2}
\usepackage{pifont}
\usepackage{graphicx}
\usepackage{enumitem}
\setlist[itemize]{nosep}
\setlist[enumerate]{nosep}
\usepackage[colorlinks=true, linkcolor=blue, citecolor=blue, urlcolor=blue]{hyperref}
\usepackage{xcolor}
\newcommand{\edit}[1]{\textcolor{black}{#1}}

\begin{document}

\title{A dynamic circuit for the honeycomb Floquet code}
\author{Jahan Claes}
\begin{abstract}
In the typical implementation of a quantum error-correcting code, each stabilizer is measured by entangling one or more ancilla qubits with the data qubits and measuring the ancilla qubits to deduce the value of the stabilizer. Recently, the dynamic circuit approach has been introduced, in which stabilizers are measured without ancilla qubits. Here, we demonstrate that dynamic circuits are particularly useful for the Floquet code. Our dynamic circuit increases the timelike distance of the code, automatically removes leakage, and both significantly increases the threshold and lowers the logical error rate compared to the standard ancilla-based circuit. At a physical error rate of $10^{-3}$, we estimate a nearly $3\times$ reduction in the number of qubits required to reach a $10^{-12}$ logical error rate.
\end{abstract}
\maketitle

\section{Introduction}

Quantum error correction (QEC) will be necessary to reach the logical error rates required for long quantum algorithms. QEC is performed by repeatedly measuring a set of Pauli operators, known as stabilizers, that detect the presence of errors~\cite{gottesmanStabilizerCodesQuantum1997,poulinStabilizerFormalismOperator2005,baconOperatorQuantumError2006,kitaev1997quantum,hastings2021dynamically}. Typically, stabilizers of a QEC code are measured using one or more ancilla qubits that are entangled with the data qubits of the code~\cite{dennis2002topological,fowler2012surface,chamberland2018flag,gidney2023new}. Recently, the idea of \textit{dynamic} or \textit{morphing} stabilizer measurement circuits has been introduced~\cite{mcewen2023relaxing,eickbusch2024demonstrating,gidney2023new,shaw2025lowering,debroy2024luci,higgott2025handling,debroy2025diamond,yoshida2025low}, in which stabilizers of the code are measured without ancilla qubits. Instead, to measure a Pauli operator $P:=\otimes_i P_i$, we run a short quantum circuit $U$ that sends $P\mapsto P_{i^*}$, where $P_{i^*}$ is a single-qubit operator which can be directly measured. This approach may use fewer qubits~\cite{shaw2025lowering,gidney2023new} or couplers~\cite{mcewen2023relaxing,debroy2024luci} or remove leakage~\cite{mcewen2023relaxing,yoshida2025low}, but it may also reduce the distance of the code~\cite{gidney2023new,yoshida2025low}.

The \edit{honeycomb} Floquet code is a QEC code defined on the honeycomb lattice, in which stabilizers are given by the product of two-qubit Pauli gauge measurements~\cite{hastings2021dynamically}. In previous work, each gauge measurement was performed using an ancilla qubit~\cite{gidney2021fault,gidney2022benchmarking}. \edit{Here,} we demonstrate that the \edit{honeycomb} Floquet code is particularly amenable to the dynamic circuit approach. We present a dynamic circuit with several notable differences compared to the standard ancilla-based circuit:
\begin{itemize}
    \setlength{\parsep}{0pt}
    \item The dynamic circuit removes leakage by measuring each qubit after four layers of two-qubit gates. It is not obvious how to do this \edit{based on} the standard circuit \edit{using modifications like} ``walking"~\cite{mcewen2023relaxing}. (\ding{51})
    \item For an $L_1\times L_2$ patch of code, the dynamic circuit uses $L_1L_2$ qubits and $\frac{3}{2}L_1L_2$ couplers, compared to $\frac{5}{2} L_1L_2$ qubits and $3L_1L_2$ couplers in the standard circuit. (\ding{51})
    \item To reach a distance $d$, the dynamic circuit requires $(L_1,L_2)=(2d,3d)$ compared to $(d,2d)$ in the standard circuit. (\ding{55})
    \item To \edit{reach} a timelike distance $d_t$, the dynamic circuit requires $d_t$ \edit{rounds of stabilizer measurements}, compared to $\frac{4}{3}d_t$ in the standard circuit. (\ding{51})
    \item \edit{Using a MWPM (correlated matching) decoder,} the threshold of the dynamic circuit is $.29\%$ \edit{($.35\%$)} under circuit-level depolarizing noise compared to \edit{$.22\%$ ($.27\%$)} in the standard circuit. (\ding{51})
    \item The dynamic circuit has a much lower logical error rate than the standard circuit, with a teraquop qubit count a factor of $2-3$ lower at $p=10^{-3}$ (\ding{51})
\end{itemize}
In the next few sections, we review the dynamic circuit approach and \edit{honeycomb} Floquet code, introduce our dynamic circuit, and explain each bullet point above. We conclude with some potential future work.

\emph{Note added}---After posting this work, it was pointed out that a dynamic circuit for the \edit{honeycomb} Floquet code was previously considered in~\cite[Appendix C]{benito2025comparative}, where they also found an increased threshold and reduced overhead despite a reduced spatial distance. Our work builds on this by providing a more complete explanation of the dynamic circuit, designing the circuit to eliminate leakage, and establishing the circuit's increased timelike distance.

\section{Previous work: Dynamic circuits in the surface code}

\begin{figure}
    \centering
    \includegraphics[width=\columnwidth]{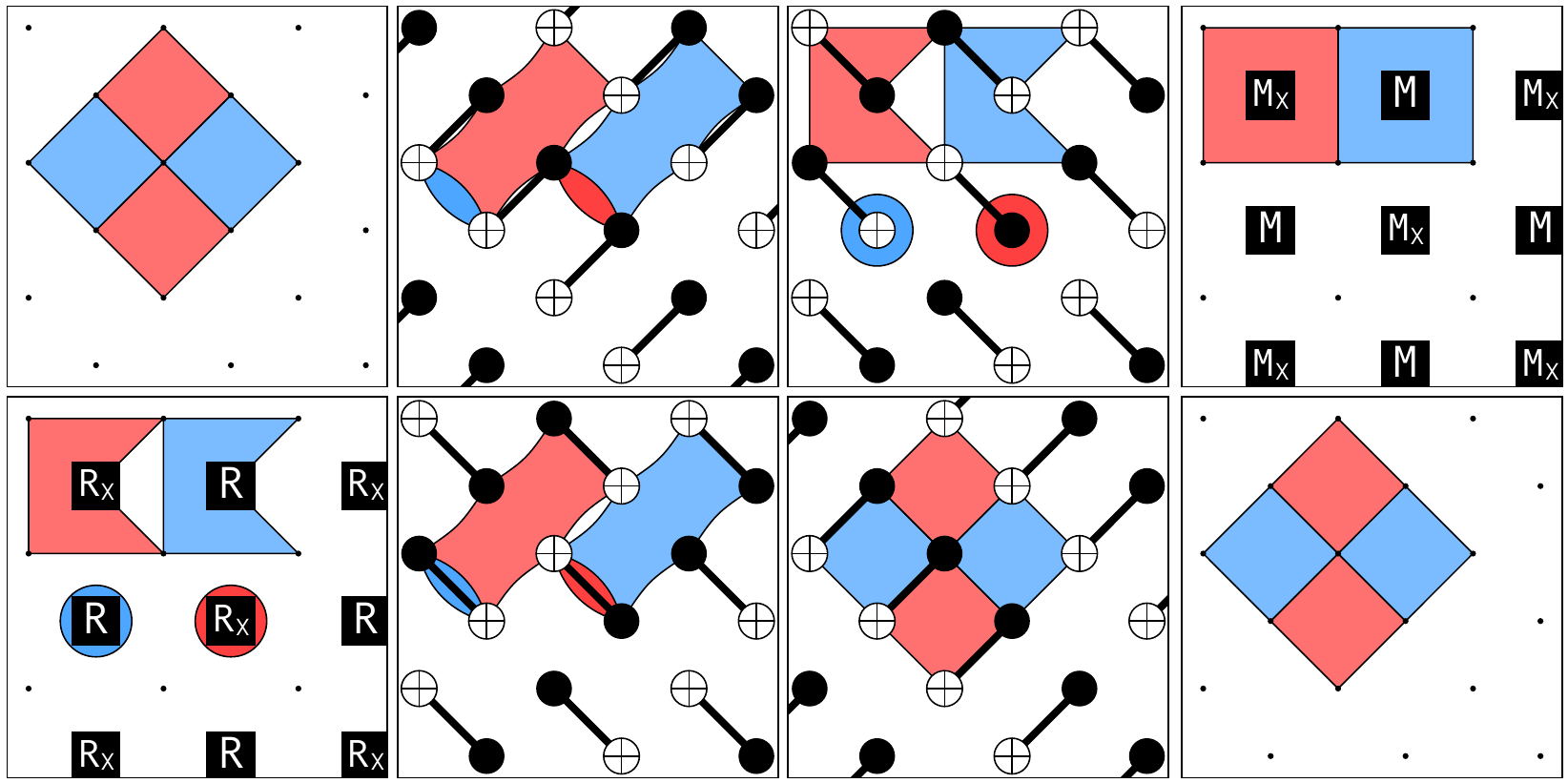}
    \caption{The dynamic circuit for measuring the stabilizers of the \edit{un}rotated surface code. \edit{Here, $M$ ($M_X$) and $R$ ($R_x$) represent measurement and reset in the $Z$ ($X$) basis.} \edit{We illustrate four neighboring stabilizers of the surface code in the first panel, with $Z$ ($X$) stabilizers in blue (red), and track their evolution through the circuit in subsequent panels}. Half the stabilizers shrink and are measured, while the neighboring stabilizers grow.}
    \label{fig:td_surface}
\end{figure}

As an introduction to dynamic circuits and their circuit-level fault distance, we review a dynamic circuit for the unrotated surface code \edit{from}~\cite{mcewen2023relaxing}, \edit{illustrated in Fig.~\ref{fig:td_surface}}. This circuit uses two layers of gates to shrink stabilizers into one-qubit Pauli operators and measure them, then \edit{applies the layers} in reverse to restore the surface code state. While half the stabilizers shrink and are measured, the other half grow. To measure \edit{the remaining} stabilizers, we re-run a shifted version of the circuit.

We consider two features of this circuit: its spatial distance $d$, and its timelike distance $d_t$. Because some stabilizers become larger as the circuit runs, the spatial distance of a dynamic circuit may be reduced, as error strings can travel faster between larger stabilizers. In addition, \edit{an elementary two-qubit gate error on a two-qubit gate $CX_{i,j}$ like $Z_iZ_j$ flips two data qubits rather than a data and ancilla qubit}. For the unrotated surface code these faster \edit{do not align with logical operators} and thus don't reduce the distance; however, if we applied this circuit to the \emph{rotated} surface code~\cite{tomita2014low}, the spatial distance would be halved.
\label{sec:TimeDynamicSurface}

As for the timelike distance, note the standard circuit measures all stabilizers in four gate layers and one measurement layer (see, e.g.,~\cite{fowler2012surface}), while the dynamic circuit only measures half the stabilizers in the same depth. Naively, one might think the dynamic circuit requires double the circuit depth to reach a timelike distance $d_t$ compared to the standard circuit. However, there is a subtle distinction between the circuits. The standard circuit compares the value of the stabilizer at time $t$ to the value at time $(t+1)$ to detect errors; the dynamic circuit compares the value of the stabilizer to the single-qubit reset operation that initialized that stabilizer. As a result, while we only measure the stabilizers half as often in the time-dynamic circuit, they effectively provide twice as much information, and we reach distance $d_t$ in the same circuit depth.

\section{A review of the honeycomb Floquet code}\label{sec:FloquetReview}

\begin{figure*}
\includegraphics[width=\textwidth]{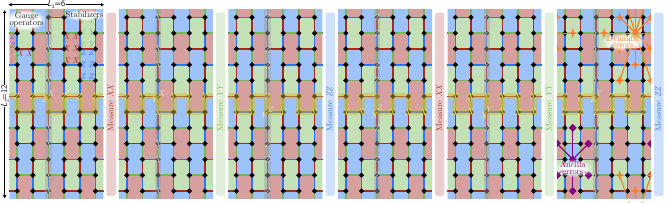}
\caption{An illustration of an $L_1\times L_2$ patch of \edit{honeycomb} Floquet code on a torus. The rightmost panel illustrates the gauge operators and stabilizers of the Floquet code. The stabilizers are supported on hexagons, with red/green/blue hexagons denoting $\otimes^6X$/$\otimes^6Y$/$\otimes^6Z$ stabilizers. Each bond supports two-qubit gauge operators that commute with all stabilizers, with red/green/blue bonds denoting $XX$/$YY$/$ZZ$ gauge operators. Rather than directly measure the stabilizers, we sequentially measure the $XX$/$YY$/$ZZ$ gauge operators, with the product of the gauge operators around each hexagon giving the value of the corresponding stabilizer. No single representative of the logical operators commutes with all gauge operators; instead, at each step, we multiply the current representative by some previous gauge operators to create a new representative that commutes with the next gauge operators. We show a horizontal (yellow) and vertical (gray) observable as they evolve through time. In the last panel, we illustrate pairs of stabilizers that can be flipped by a single physical error for the standard (purple) and dynamic (orange) circuits. In the standard circuit, errors flip neighboring pairs of stabilizers, while in the dynamic circuit they may flip next-nearest-neighbors.
}
\label{fig:FloquetExplain}
\end{figure*}

The \edit{honeycomb} Floquet code~\cite{hastings2021dynamically} is illustrated in Fig.~\ref{fig:FloquetExplain}. Qubits are arranged at the vertices of a periodic honeycomb lattice with horizontal and vertical dimensions $L_1$ and $L_2$. Each hexagon corresponds to a six-qubit stabilizer, given by $\otimes^6X$/$\otimes^6Y$/$\otimes^6 Z$ for red/green/blue hexagons.

Rather than directly measuring the stabilizers of the Floquet code, we infer their value by measuring gauge operators that commute with all stabilizers. Each bond corresponds to a two-qubit gauge measurement, given by $XX$/$YY$/$ZZ$ for red/green/blue bonds. We repeatedly measure \edit{$XX\rightarrow YY\rightarrow ZZ\rightarrow\cdots$} gauge operators in sequence. The product of the gauge operators around a cell then gives the value of that stabilizer. This allows us to deduce the stabilizer values without directly performing six-qubit measurements.

A subtlety arises when considering logical operators in the Floquet code. For any logical degree of freedom, no fixed representative of the logical operator commutes with all gauge measurements. Instead, we modify the logical representative at each time step, multiplying it by gauge operators that have just been measured to form a new representative that commutes with the next gauge measurement. We illustrate this process for a horizontal/vertical logical operator pair in~\ref{fig:FloquetExplain}. While the measurement cycle has period $3$, the logical representatives have period $6$.

In this paper, we consider the Floquet code on a torus \edit{to avoid the additional complexity of boundaries}; however, it is also possible to define planar Floquet codes \edit{with boundaries}~\cite{gidney2022benchmarking,haah2022boundaries} \edit{which can be straightforwardly realized with our dynamic circuits}. Since our code is on a torus, it encodes two logical qubits; however, we restrict our attention to a single logical qubit as only one logical qubit survives introducing boundaries.

\section{Standard vs dynamic circuit for the Floquet code}\label{sec:TimeDynamicHoneycomb}

\begin{figure*}
    \includegraphics[width=\textwidth]{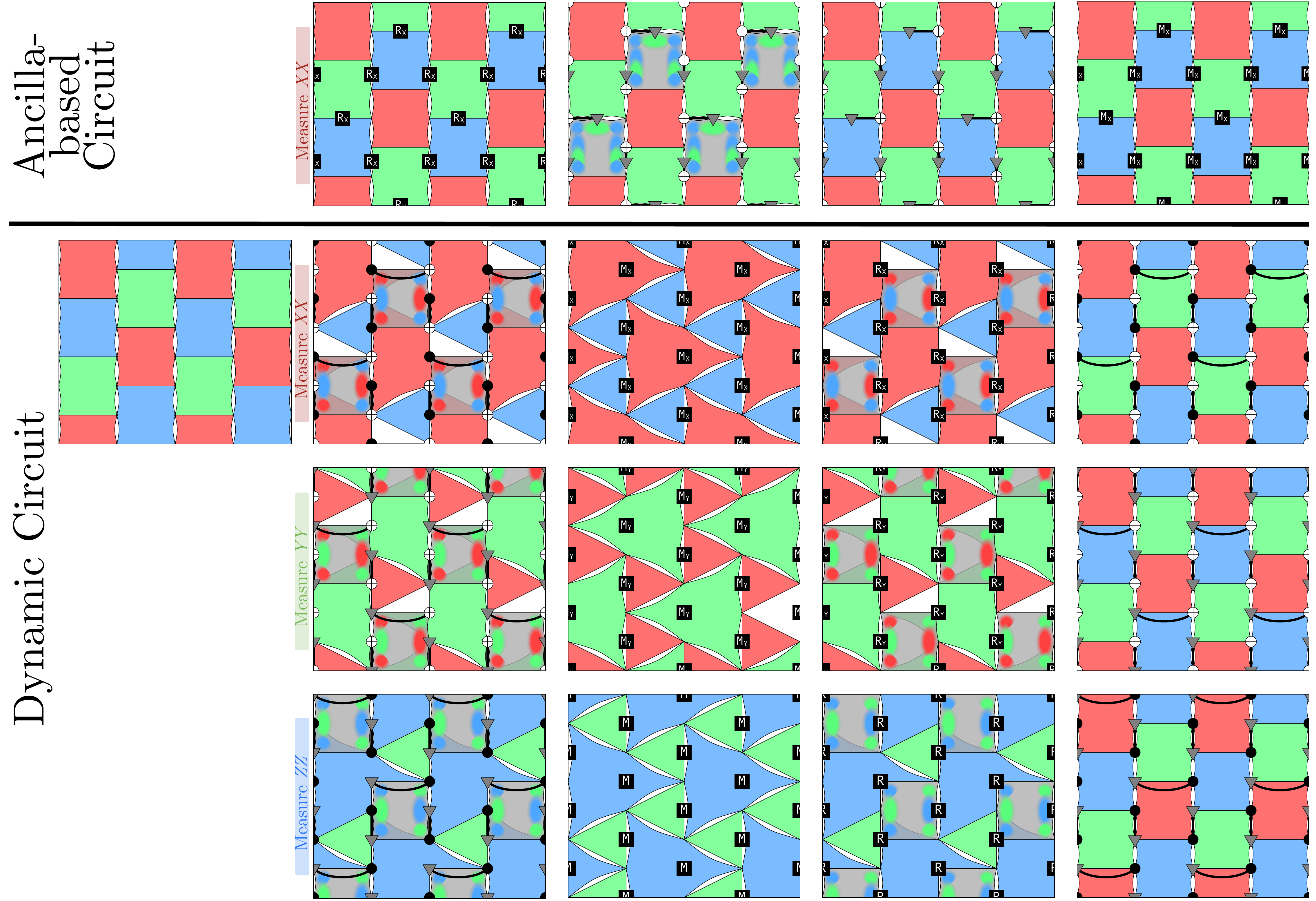}
    \caption{Top: Standard circuit for performing $XX$ gauge measurements, using one ancilla per gauge operator. The circuits for $YY$ and $ZZ$ are similar. Bottom: Gauge measurements in the dynamic circuit. We measure $XX$ gauge operators by shrinking them to a single qubit, which we then measure. In the $YY$ and $ZZ$ gauge measurements we alternate the measured qubits to remove leakage. In the next three gauge measurements (not shown), we reverse the target and control of each gate and continue the alternating measurement pattern. The two circuits have the same number of two-qubit gates, resets, and measurements.}
    \label{fig:circuits}
\end{figure*}

We present two circuits for measuring the gauge operators, the standard ancilla-based circuit and our new dynamic one, illustrated in Fig.~\ref{fig:circuits}. The standard circuit requires one qubit per gauge operator, while the dynamic one requires no ancilla qubits and operates by shrinking the gauge operator onto a single qubit and measuring that qubit directly. Note that in the surface code, switching to the dynamic circuit meant we could only measure half as many operators at once (but got twice as much information per measurement). In contrast, in the Floquet code we get improved measurements while measuring all $XX$ operators simultaneously, which is the source of improved performance.

\edit{We have designed our dynamic circuit so that the gauge operators are shrunk onto alternating qubits in each measurement round. Thus, after two measurement rounds, every qubit has been measured and reset, allowing for the removal of leakage that has occured in the previous two rounds}.

We note that individual errors in the standard circuit only flip neighboring stabilizers, as can be seen from the evolution of the stabilizers in the top of Fig.~\ref{fig:circuits}. However, the dynamic circuit causes next-nearest-neighbor stabilizers to touch (see the middle columns of Fig.~\ref{fig:circuits}), so individual errors can flip next-nearest-neighbors. This is illustrated in the last panel of Fig.~\ref{fig:FloquetExplain}, from which it follows that the standard (dynamic) circuit requires $(L_1,L_2)=(d,2d)$ ($(L_1,L_2)=(2d,3d)$)~\footnote{Note that an earlier paper~\cite{gidney2021fault} misstated the vertical circuit distance, as was noted in a later paper by some of the same authors~\cite{gidney2022benchmarking}}. \edit{We have verified the resulting $d=6,12,18$ circuits have the correct distance against graphlike errors using STIM's \texttt{shortest\_graphlike\_error()}. We have also verified that the $d=6$ circuits have the correct distance against more general errors using STIM's heuristic \texttt{search\_for\_undetectable\_logical\_errors()}}. The qubit number for a distance-$d$ code is then $5d^2$ for the standard circuit, vs $6d^2$ for the dynamic circuit.

To understand the logical error rate and spacetime overhead per logical operation, we must also consider the timelike distance $d_t$, since a braiding~\cite{raussendorf2007fault,raussendorf2007topological,fowler2009high,fowler2012surface} or lattice surgery~\cite{horsman2012surface,litinski2018lattice,fowler2018low} operation requires $d_t\approx d$. The timelike distance of our dynamic circuit is \textit{increased} due to the improved measurements; the standard circuit requires measuring $\{XX,YY,ZZ\}$ $4d_t/3$ times, while the dynamic circuit requires only $d_t$. \edit{We explain the timelike distance, which appears to have been neglected in previous work on Floquet codes, in more detail in Appendix~\ref{appendix:timelike}.}

Our dynamic circuit thus requires fewer measurement \edit{cycles} per logical operation. However, every layer of our dynamic circuit must be executed sequentially, meaning a full cycle takes time $(6t_\text{gate}+3t_\text{meas}+3t_\text{reset})$. This contrasts with the standard circuit, in which the first layer of gates to measure $YY$ commutes with the last layer of gates to measure $XX$, and measured qubits do not need to be immediately reset. The standard circuit can therefore be compressed; in~\cite{gidney2021fault} they provide a circuit that takes amortized time $(3t_\text{gate}+3\max[t_\text{meas},t_\text{reset}])$ provided the hardware can execute measure and reset operations in parallel. As a result, depending on hardware constraints, a logical operation in the dynamic circuit may still be slower.

For our simulations, we will compare the logical error rates of the dynamic and standard circuits at the same $d=d_t$, noting that the dynamic circuit requires slightly more qubits and the standard circuit requires slightly more measurements.

\section{Numerical results}\label{sec:Numerics}

\begin{figure}[h]
    \includegraphics[width=\columnwidth]{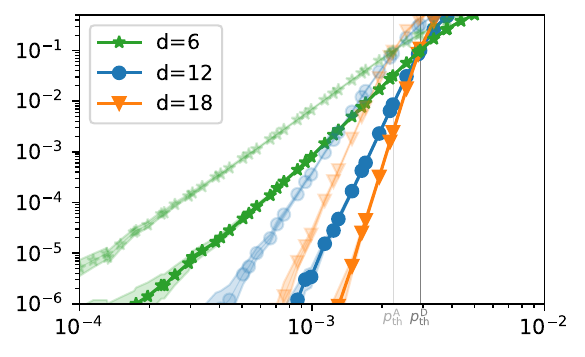}
    \caption{Logical error rates \edit{for a single pair of logical observables} in the dynamic circuit (bold) and standard circuit (faded) for various circuit-level distances, using MWPM. Estimated thresholds are shown with horizontal lines. The dynamic circuit demonstrates significant improvement in both threshold and logical error rate. Circuits simulated with STIM~\cite{gidney2021stim} and decoded with PyMatching~\cite{higgott2025sparse}.}
    \label{fig:ResultsLER}
\end{figure}
Fig.~\ref{fig:ResultsLER} gives our simulation results using a standard depolarizing noise model (see Appendix~\ref{appendix:noise_model})\edit{decoded using MWPM. We include correlated matching results in Appendix~\ref{appendix:numerics}.} \edit{The logical error rate is given for a single pair of horizontal/vertical operators}. The dynamic circuit has a threshold of $\approx .29\%$ \edit{($.35\%$)}, compared to \edit{$\approx .22\%$ ($.27\%$)} for the standard circuit \edit{using MWPM (correlated matching)}. Related to the higher threshold, the logical error rate of the distance-$d$ dynamic circuit has a significantly lower error rate; approximately $6\times$ lower at $d=6$, $50\times$ lower at $d=12$, and $400\times$ lower at $d=18$. For each curve, the exponent of the logical error rate is slightly larger than $\lceil \frac{d}{2}\rceil$; this indicates that there are higher-distance errors that contribute significantly to the logical error rate.

A common figure of merit for comparing codes is the \textit{teraquop \edit{overhead}}~\cite{gidney2021fault}, the number of qubits needed to incur a $10^{-12}$ error probability \edit{per} timelike distance $d_t=d$. \edit{Extrapolating to larger $d$ by assuming the the logical error rate decays exponentially}, for $p=10^{-3}$ we require \edit{$d=29$ ($d=23$)} for the dynamic circuit versus $d=53$ \edit{($d=39$)} for the standard circuit \edit{under MWPM (correlated matching)}. This corresponds to \edit{5000 (3200)} qubits for the dynamic circuit and \edit{14000 (7600)} qubits for the standard circuit. 

\section{Conclusion}\label{sec:Conclusion}

We have demonstrated that our dynamic circuit increases the threshold and lowers the resource overhead of the \edit{honeycomb} Floquet code. Our work leaves open a few avenues for future exploration.

First, we have neglected to introduce boundaries in our circuits~\cite{gidney2022benchmarking,haah2022boundaries}, although we do not expect boundaries to significantly affect our results. We also have not simulated a stability experiment~\cite{gidney2022stability} to estimate the timelike error rates. We also only simulated a simple depolarizing error model; the advantage of the dynamic circuit will depend on the error model, and become larger for models that include leakage.

Finally, numerous generalizations of the \edit{honeycomb} Floquet code have been introduced that also reduce stabilizer measurements to a series of two-qubit gauge measurements~\cite{paetznick2023performance,townsend2023floquetifying,davydova2023floquet,rodatz2024floquetifying,zhang2022x,dua2024engineering,de2024xyz,higgott2024constructions,fahimniya2023fault,alam2024dynamical,setiawan2024tailoring}; it is an open question if these codes could also benefit from our dynamic approach. In particular, the square-octagon Floquet code~\cite{paetznick2023performance} will not have a reduced spatial distance in the dynamic circuit, as the gauge operators do not lie along the logical operators; a circuit-level simulation of this code will likely show dramatic improvements.

\begin{acknowledgments}
The ancilla-based simulations were performed using code included with~\cite{gidney2021fault}. The dynamic simulations also made substantial use of tools included in that code base. 

\ 

\emph{Data availability}--Stim circuits and simulation data are available at~\cite{zenodo}.
\end{acknowledgments}

\appendix

\section{Noise Model}
\label{appendix:noise_model}
To be consistent with previous work~\cite{gidney2021fault,gidney2022benchmarking}, we use the standard depolarizing noise model:
\begin{itemize}
    \item Single-qubit gates are followed by a single-qubit depolarizing channel with strength $p$.
    \item Two-qubit gates are followed by a two-qubit depolarizing channel with strength $p$.
    \item Idle qubits during single or two-qubit gates incur a single-qubit depolarizing channel with strength $p$.
    \item Initialization in $|0\rangle$ gets flipped with probability $p$, and similar for $|+\rangle$.
    \item Measurement results are flipped with probability $p$.
\end{itemize}
Note that idle errors are significantly more damaging to the dynamic circuit, because once the standard circuit is compacted it has few idle locations. If idle errors were negligible, the advantage of the dynamic circuit would increase.

\section{Timelike distance}
\label{appendix:timelike}

\edit{In the standard circuit, the $\otimes^6Z$ detector is formed as the product of gauge operators at times $XX(t)$, $YY(t)$, $XX(t+1)$, and $YY(t+1)$. Notably, this means that a detector spans \textit{four} gauge measurements, and has a timelike extent of $4/3$ of a full round. In contrast, in the dynamic circuit, this detector is formed by multiplying $ZZ(t)$ and $YY(t+1)$ and has timelike extent of one full round (neither of these statements are obvious, and can best be understood by examining the circuit} \href{https://algassert.com/crumble#circuit=Q(1,0)0;Q(1,1)1;Q(1,2)2;Q(1,3)3;Q(1,4)4;Q(1,5)5;Q(1,6)6;Q(1,7)7;Q(1,8)8;Q(1,9)9;Q(1,10)10;Q(1,11)11;Q(3,0)12;Q(3,1)13;Q(3,2)14;Q(3,3)15;Q(3,4)16;Q(3,5)17;Q(3,6)18;Q(3,7)19;Q(3,8)20;Q(3,9)21;Q(3,10)22;Q(3,11)23;Q(5,0)24;Q(5,1)25;Q(5,2)26;Q(5,3)27;Q(5,4)28;Q(5,5)29;Q(5,6)30;Q(5,7)31;Q(5,8)32;Q(5,9)33;Q(5,10)34;Q(5,11)35;Q(7,0)36;Q(7,1)37;Q(7,2)38;Q(7,3)39;Q(7,4)40;Q(7,5)41;Q(7,6)42;Q(7,7)43;Q(7,8)44;Q(7,9)45;Q(7,10)46;Q(7,11)47;Q(9,0)48;Q(9,1)49;Q(9,2)50;Q(9,3)51;Q(9,4)52;Q(9,5)53;Q(9,6)54;Q(9,7)55;Q(9,8)56;Q(9,9)57;Q(9,10)58;Q(9,11)59;Q(11,0)60;Q(11,1)61;Q(11,2)62;Q(11,3)63;Q(11,4)64;Q(11,5)65;Q(11,6)66;Q(11,7)67;Q(11,8)68;Q(11,9)69;Q(11,10)70;Q(11,11)71;Q(13,0)72;Q(13,1)73;Q(13,2)74;Q(13,3)75;Q(13,4)76;Q(13,5)77;Q(13,6)78;Q(13,7)79;Q(13,8)80;Q(13,9)81;Q(13,10)82;Q(13,11)83;Q(15,0)84;Q(15,1)85;Q(15,2)86;Q(15,3)87;Q(15,4)88;Q(15,5)89;Q(15,6)90;Q(15,7)91;Q(15,8)92;Q(15,9)93;Q(15,10)94;Q(15,11)95;CX_87_3_93_9_2_1_4_5_8_7_10_11_0_12_6_18_13_14_17_16_19_20_23_22_15_27_21_33_26_25_28_29_32_31_34_35_24_36_30_42_37_38_41_40_43_44_47_46_39_51_45_57_50_49_52_53_56_55_58_59_48_60_54_66_61_62_65_64_67_68_71_70_63_75_69_81_74_73_76_77_80_79_82_83_72_84_78_90_85_86_89_88_91_92_95_94;TICK;MX_87_93_2_4_8_10_0_6_13_17_19_23_15_21_26_28_32_34_24_30_37_41_43_47_39_45_50_52_56_58_48_54_61_65_67_71_63_69_74_76_80_82_72_78_85_89_91_95;TICK;RX_87_93_2_4_8_10_0_6_13_17_19_23_15_21_26_28_32_34_24_30_37_41_43_47_39_45_50_52_56_58_48_54_61_65_67_71_63_69_74_76_80_82_72_78_85_89_91_95;MARKX(0)28_41_30;TICK;CX_87_3_93_9_2_1_4_5_8_7_10_11_0_12_6_18_13_14_17_16_19_20_23_22_15_27_21_33_26_25_28_29_32_31_34_35_24_36_30_42_37_38_41_40_43_44_47_46_39_51_45_57_50_49_52_53_56_55_58_59_48_60_54_66_61_62_65_64_67_68_71_70_63_75_69_81_74_73_76_77_80_79_82_83_72_84_78_90_85_86_89_88_91_92_95_94;TICK;XCY_1_85_7_91_3_2_5_6_9_8_11_0_16_4_22_10_14_15_18_17_20_21_12_23_25_13_31_19_27_26_29_30_33_32_35_24_40_28_46_34_38_39_42_41_44_45_36_47_49_37_55_43_51_50_53_54_57_56_59_48_64_52_70_58_62_63_66_65_68_69_60_71_73_61_79_67_75_74_77_78_81_80_83_72_88_76_94_82_86_87_90_89_92_93_84_95;TICK;MY_3_9_1_5_7_11_12_18_14_16_20_22_27_33_25_29_31_35_36_42_38_40_44_46_51_57_49_53_55_59_60_66_62_64_68_70_75_81_73_77_79_83_84_90_86_88_92_94;TICK;RY_3_9_1_5_7_11_12_18_14_16_20_22_27_33_25_29_31_35_36_42_38_40_44_46_51_57_49_53_55_59_60_66_62_64_68_70_75_81_73_77_79_83_84_90_86_88_92_94;MARKY(0)29_40_42;TICK;XCY_1_85_7_91_3_2_5_6_9_8_11_0_16_4_22_10_14_15_18_17_20_21_12_23_25_13_31_19_27_26_29_30_33_32_35_24_40_28_46_34_38_39_42_41_44_45_36_47_49_37_55_43_51_50_53_54_57_56_59_48_64_52_70_58_62_63_66_65_68_69_60_71_73_61_79_67_75_74_77_78_81_80_83_72_88_76_94_82_86_87_90_89_92_93_84_95;TICK;CY_5_89_11_95_1_0_3_4_7_6_9_10_14_2_20_8_12_13_16_15_18_19_22_21_29_17_35_23_25_24_27_28_31_30_33_34_38_26_44_32_36_37_40_39_42_43_46_45_53_41_59_47_49_48_51_52_55_54_57_58_62_50_68_56_60_61_64_63_66_67_70_69_77_65_83_71_73_72_75_76_79_78_81_82_86_74_92_80_84_85_88_87_90_91_94_93;TICK;M_87_93_2_4_8_10_0_6_13_17_19_23_15_21_26_28_32_34_24_30_37_41_43_47_39_45_50_52_56_58_48_54_61_65_67_71_63_69_74_76_80_82_72_78_85_89_91_95;MARKZ(0)28_41_30;TICK;R_87_93_2_4_8_10_0_6_13_17_19_23_15_21_26_28_32_34_24_30_37_41_43_47_39_45_50_52_56_58_48_54_61_65_67_71_63_69_74_76_80_82_72_78_85_89_91_95;MARKZ(0)28_30_41;TICK;CY_5_89_11_95_1_0_3_4_7_6_9_10_14_2_20_8_12_13_16_15_18_19_22_21_29_17_35_23_25_24_27_28_31_30_33_34_38_26_44_32_36_37_40_39_42_43_46_45_53_41_59_47_49_48_51_52_55_54_57_58_62_50_68_56_60_61_64_63_66_67_70_69_77_65_83_71_73_72_75_76_79_78_81_82_86_74_92_80_84_85_88_87_90_91_94_93;TICK;CX_3_87_9_93_1_2_5_4_7_8_11_10_12_0_18_6_14_13_16_17_20_19_22_23_27_15_33_21_25_26_29_28_31_32_35_34_36_24_42_30_38_37_40_41_44_43_46_47_51_39_57_45_49_50_53_52_55_56_59_58_60_48_66_54_62_61_64_65_68_67_70_71_75_63_81_69_73_74_77_76_79_80_83_82_84_72_90_78_86_85_88_89_92_91_94_95;TICK;MX_3_9_1_5_7_11_12_18_14_16_20_22_27_33_25_29_31_35_36_42_38_40_44_46_51_57_49_53_55_59_60_66_62_64_68_70_75_81_73_77_79_83_84_90_86_88_92_94;TICK;RX_3_9_1_5_7_11_12_18_14_16_20_22_27_33_25_29_31_35_36_42_38_40_44_46_51_57_49_53_55_59_60_66_62_64_68_70_75_81_73_77_79_83_84_90_86_88_92_94;MARKX(0)40_29_42;TICK;CX_3_87_9_93_1_2_5_4_7_8_11_10_12_0_18_6_14_13_16_17_20_19_22_23_27_15_33_21_25_26_29_28_31_32_35_34_36_24_42_30_38_37_40_41_44_43_46_47_51_39_57_45_49_50_53_52_55_56_59_58_60_48_66_54_62_61_64_65_68_67_70_71_75_63_81_69_73_74_77_76_79_80_83_82_84_72_90_78_86_85_88_89_92_91_94_95;TICK;XCY_85_1_91_7_2_3_6_5_8_9_0_11_4_16_10_22_15_14_17_18_21_20_23_12_13_25_19_31_26_27_30_29_32_33_24_35_28_40_34_46_39_38_41_42_45_44_47_36_37_49_43_55_50_51_54_53_56_57_48_59_52_64_58_70_63_62_65_66_69_68_71_60_61_73_67_79_74_75_78_77_80_81_72_83_76_88_82_94_87_86_89_90_93_92_95_84;TICK;MY_87_93_2_4_8_10_0_6_13_17_19_23_15_21_26_28_32_34_24_30_37_41_43_47_39_45_50_52_56_58_48_54_61_65_67_71_63_69_74_76_80_82_72_78_85_89_91_95;MARKY(0)28_41_30;DT(3,5,0)rec[-39]_rec[-41]_rec[-45]_rec[-135]_rec[-137]_rec[-141];DT(3,11,0)rec[-37]_rec[-42]_rec[-43]_rec[-133]_rec[-138]_rec[-139];DT(5,2,0)rec[-34]_rec[-36]_rec[-40]_rec[-130]_rec[-132]_rec[-136];DT(5,8,0)rec[-32]_rec[-35]_rec[-38]_rec[-128]_rec[-131]_rec[-134];DT(7,5,0)rec[-27]_rec[-29]_rec[-33]_rec[-123]_rec[-125]_rec[-129];DT(7,11,0)rec[-25]_rec[-30]_rec[-31]_rec[-121]_rec[-126]_rec[-127];DT(9,2,0)rec[-22]_rec[-24]_rec[-28]_rec[-118]_rec[-120]_rec[-124];DT(9,8,0)rec[-20]_rec[-23]_rec[-26]_rec[-116]_rec[-119]_rec[-122];DT(11,5,0)rec[-15]_rec[-17]_rec[-21]_rec[-111]_rec[-113]_rec[-117];DT(11,11,0)rec[-13]_rec[-18]_rec[-19]_rec[-109]_rec[-114]_rec[-115];DT(13,2,0)rec[-10]_rec[-12]_rec[-16]_rec[-106]_rec[-108]_rec[-112];DT(13,8,0)rec[-8]_rec[-11]_rec[-14]_rec[-104]_rec[-107]_rec[-110];DT(15,1,0)rec[-4]_rec[-46]_rec[-48]_rec[-100]_rec[-142]_rec[-144];DT(15,5,0)rec[-3]_rec[-5]_rec[-9]_rec[-99]_rec[-101]_rec[-105];DT(15,7,0)rec[-2]_rec[-44]_rec[-47]_rec[-98]_rec[-140]_rec[-143];DT(15,11,0)rec[-1]_rec[-6]_rec[-7]_rec[-97]_rec[-102]_rec[-103];TICK;RY_87_93_2_4_8_10_0_6_13_17_19_23_15_21_26_28_32_34_24_30_37_41_43_47_39_45_50_52_56_58_48_54_61_65_67_71_63_69_74_76_80_82_72_78_85_89_91_95;TICK;XCY_85_1_91_7_2_3_6_5_8_9_0_11_4_16_10_22_15_14_17_18_21_20_23_12_13_25_19_31_26_27_30_29_32_33_24_35_28_40_34_46_39_38_41_42_45_44_47_36_37_49_43_55_50_51_54_53_56_57_48_59_52_64_58_70_63_62_65_66_69_68_71_60_61_73_67_79_74_75_78_77_80_81_72_83_76_88_82_94_87_86_89_90_93_92_95_84;TICK;CY_89_5_95_11_0_1_4_3_6_7_10_9_2_14_8_20_13_12_15_16_19_18_21_22_17_29_23_35_24_25_28_27_30_31_34_33_26_38_32_44_37_36_39_40_43_42_45_46_41_53_47_59_48_49_52_51_54_55_58_57_50_62_56_68_61_60_63_64_67_66_69_70_65_77_71_83_72_73_76_75_78_79_82_81_74_86_80_92_85_84_87_88_91_90_93_94;TICK;M_3_9_1_5_7_11_12_18_14_16_20_22_27_33_25_29_31_35_36_42_38_40_44_46_51_57_49_53_55_59_60_66_62_64_68_70_75_81_73_77_79_83_84_90_86_88_92_94;DT(3,4,1)rec[-39]_rec[-40]_rec[-48]_rec[-135]_rec[-136]_rec[-144];DT(3,10,1)rec[-37]_rec[-38]_rec[-47]_rec[-133]_rec[-134]_rec[-143];DT(5,7,1)rec[-32]_rec[-33]_rec[-41]_rec[-128]_rec[-129]_rec[-137];DT(5,11,1)rec[-31]_rec[-34]_rec[-42]_rec[-127]_rec[-130]_rec[-138];DT(7,4,1)rec[-27]_rec[-28]_rec[-36]_rec[-123]_rec[-124]_rec[-132];DT(7,10,1)rec[-25]_rec[-26]_rec[-35]_rec[-121]_rec[-122]_rec[-131];DT(9,7,1)rec[-20]_rec[-21]_rec[-29]_rec[-116]_rec[-117]_rec[-125];DT(9,11,1)rec[-19]_rec[-22]_rec[-30]_rec[-115]_rec[-118]_rec[-126];DT(11,4,1)rec[-15]_rec[-16]_rec[-24]_rec[-111]_rec[-112]_rec[-120];DT(11,10,1)rec[-13]_rec[-14]_rec[-23]_rec[-109]_rec[-110]_rec[-119];DT(13,7,1)rec[-8]_rec[-9]_rec[-17]_rec[-104]_rec[-105]_rec[-113];DT(13,11,1)rec[-7]_rec[-10]_rec[-18]_rec[-103]_rec[-106]_rec[-114];DT(15,0,1)rec[-6]_rec[-43]_rec[-46]_rec[-102]_rec[-139]_rec[-142];DT(15,6,1)rec[-5]_rec[-44]_rec[-45]_rec[-101]_rec[-140]_rec[-141];DT(15,4,1)rec[-3]_rec[-4]_rec[-12]_rec[-99]_rec[-100]_rec[-108];DT(15,10,1)rec[-1]_rec[-2]_rec[-11]_rec[-97]_rec[-98]_rec[-107];TICK;R_3_9_1_5_7_11_12_18_14_16_20_22_27_33_25_29_31_35_36_42_38_40_44_46_51_57_49_53_55_59_60_66_62_64_68_70_75_81_73_77_79_83_84_90_86_88_92_94;TICK;CY_89_5_95_11_0_1_4_3_6_7_10_9_2_14_8_20_13_12_15_16_19_18_21_22_17_29_23_35_24_25_28_27_30_31_34_33_26_38_32_44_37_36_39_40_43_42_45_46_41_53_47_59_48_49_52_51_54_55_58_57_50_62_56_68_61_60_63_64_67_66_69_70_65_77_71_83_72_73_76_75_78_79_82_81_74_86_80_92_85_84_87_88_91_90_93_94;TICK;CX_87_3_93_9_2_1_4_5_8_7_10_11_0_12_6_18_13_14_17_16_19_20_23_22_15_27_21_33_26_25_28_29_32_31_34_35_24_36_30_42_37_38_41_40_43_44_47_46_39_51_45_57_50_49_52_53_56_55_58_59_48_60_54_66_61_62_65_64_67_68_71_70_63_75_69_81_74_73_76_77_80_79_82_83_72_84_78_90_85_86_89_88_91_92_95_94;TICK;MX_87_93_2_4_8_10_0_6_13_17_19_23_15_21_26_28_32_34_24_30_37_41_43_47_39_45_50_52_56_58_48_54_61_65_67_71_63_69_74_76_80_82_72_78_85_89_91_95;DT(3,1,2)rec[-40]_rec[-42]_rec[-46]_rec[-136]_rec[-138]_rec[-142];DT(3,7,2)rec[-38]_rec[-41]_rec[-44]_rec[-134]_rec[-137]_rec[-140];DT(5,4,2)rec[-33]_rec[-36]_rec[-39]_rec[-129]_rec[-132]_rec[-135];DT(5,10,2)rec[-31]_rec[-35]_rec[-37]_rec[-127]_rec[-131]_rec[-133];DT(7,1,2)rec[-28]_rec[-30]_rec[-34]_rec[-124]_rec[-126]_rec[-130];DT(7,7,2)rec[-26]_rec[-29]_rec[-32]_rec[-122]_rec[-125]_rec[-128];DT(9,4,2)rec[-21]_rec[-24]_rec[-27]_rec[-117]_rec[-120]_rec[-123];DT(9,10,2)rec[-19]_rec[-23]_rec[-25]_rec[-115]_rec[-119]_rec[-121];DT(11,1,2)rec[-16]_rec[-18]_rec[-22]_rec[-112]_rec[-114]_rec[-118];DT(11,7,2)rec[-14]_rec[-17]_rec[-20]_rec[-110]_rec[-113]_rec[-116];DT(13,4,2)rec[-9]_rec[-12]_rec[-15]_rec[-105]_rec[-108]_rec[-111];DT(13,10,2)rec[-7]_rec[-11]_rec[-13]_rec[-103]_rec[-107]_rec[-109];DT(15,1,2)rec[-4]_rec[-6]_rec[-10]_rec[-100]_rec[-102]_rec[-106];DT(15,5,2)rec[-3]_rec[-45]_rec[-48]_rec[-99]_rec[-141]_rec[-144];DT(15,7,2)rec[-2]_rec[-5]_rec[-8]_rec[-98]_rec[-101]_rec[-104];DT(15,11,2)rec[-1]_rec[-43]_rec[-47]_rec[-97]_rec[-139]_rec[-143];TICK;RX_87_93_2_4_8_10_0_6_13_17_19_23_15_21_26_28_32_34_24_30_37_41_43_47_39_45_50_52_56_58_48_54_61_65_67_71_63_69_74_76_80_82_72_78_85_89_91_95;TICK;CX_87_3_93_9_2_1_4_5_8_7_10_11_0_12_6_18_13_14_17_16_19_20_23_22_15_27_21_33_26_25_28_29_32_31_34_35_24_36_30_42_37_38_41_40_43_44_47_46_39_51_45_57_50_49_52_53_56_55_58_59_48_60_54_66_61_62_65_64_67_68_71_70_63_75_69_81_74_73_76_77_80_79_82_83_72_84_78_90_85_86_89_88_91_92_95_94;TICK;XCY_1_85_7_91_3_2_5_6_9_8_11_0_16_4_22_10_14_15_18_17_20_21_12_23_25_13_31_19_27_26_29_30_33_32_35_24_40_28_46_34_38_39_42_41_44_45_36_47_49_37_55_43_51_50_53_54_57_56_59_48_64_52_70_58_62_63_66_65_68_69_60_71_73_61_79_67_75_74_77_78_81_80_83_72_88_76_94_82_86_87_90_89_92_93_84_95;TICK;MY_3_9_1_5_7_11_12_18_14_16_20_22_27_33_25_29_31_35_36_42_38_40_44_46_51_57_49_53_55_59_60_66_62_64_68_70_75_81_73_77_79_83_84_90_86_88_92_94;DT(3,4,3)rec[-39]_rec[-41]_rec[-45]_rec[-135]_rec[-137]_rec[-141];DT(3,10,3)rec[-37]_rec[-42]_rec[-43]_rec[-133]_rec[-138]_rec[-139];DT(5,1,3)rec[-34]_rec[-36]_rec[-40]_rec[-130]_rec[-132]_rec[-136];DT(5,7,3)rec[-32]_rec[-35]_rec[-38]_rec[-128]_rec[-131]_rec[-134];DT(7,4,3)rec[-27]_rec[-29]_rec[-33]_rec[-123]_rec[-125]_rec[-129];DT(7,10,3)rec[-25]_rec[-30]_rec[-31]_rec[-121]_rec[-126]_rec[-127];DT(9,1,3)rec[-22]_rec[-24]_rec[-28]_rec[-118]_rec[-120]_rec[-124];DT(9,7,3)rec[-20]_rec[-23]_rec[-26]_rec[-116]_rec[-119]_rec[-122];DT(11,4,3)rec[-15]_rec[-17]_rec[-21]_rec[-111]_rec[-113]_rec[-117];DT(11,10,3)rec[-13]_rec[-18]_rec[-19]_rec[-109]_rec[-114]_rec[-115];DT(13,1,3)rec[-10]_rec[-12]_rec[-16]_rec[-106]_rec[-108]_rec[-112];DT(13,7,3)rec[-8]_rec[-11]_rec[-14]_rec[-104]_rec[-107]_rec[-110];DT(15,2,3)rec[-4]_rec[-46]_rec[-48]_rec[-100]_rec[-142]_rec[-144];DT(15,4,3)rec[-3]_rec[-5]_rec[-9]_rec[-99]_rec[-101]_rec[-105];DT(15,8,3)rec[-2]_rec[-44]_rec[-47]_rec[-98]_rec[-140]_rec[-143];DT(15,10,3)rec[-1]_rec[-6]_rec[-7]_rec[-97]_rec[-102]_rec[-103];TICK;RY_3_9_1_5_7_11_12_18_14_16_20_22_27_33_25_29_31_35_36_42_38_40_44_46_51_57_49_53_55_59_60_66_62_64_68_70_75_81_73_77_79_83_84_90_86_88_92_94;TICK;XCY_1_85_7_91_3_2_5_6_9_8_11_0_16_4_22_10_14_15_18_17_20_21_12_23_25_13_31_19_27_26_29_30_33_32_35_24_40_28_46_34_38_39_42_41_44_45_36_47_49_37_55_43_51_50_53_54_57_56_59_48_64_52_70_58_62_63_66_65_68_69_60_71_73_61_79_67_75_74_77_78_81_80_83_72_88_76_94_82_86_87_90_89_92_93_84_95;TICK;CY_5_89_11_95_1_0_3_4_7_6_9_10_14_2_20_8_12_13_16_15_18_19_22_21_29_17_35_23_25_24_27_28_31_30_33_34_38_26_44_32_36_37_40_39_42_43_46_45_53_41_59_47_49_48_51_52_55_54_57_58_62_50_68_56_60_61_64_63_66_67_70_69_77_65_83_71_73_72_75_76_79_78_81_82_86_74_92_80_84_85_88_87_90_91_94_93;TICK;M_87_93_2_4_8_10_0_6_13_17_19_23_15_21_26_28_32_34_24_30_37_41_43_47_39_45_50_52_56_58_48_54_61_65_67_71_63_69_74_76_80_82_72_78_85_89_91_95;DT(3,3,4)rec[-36]_rec[-45]_rec[-46]_rec[-132]_rec[-141]_rec[-142];DT(3,9,4)rec[-35]_rec[-43]_rec[-44]_rec[-131]_rec[-139]_rec[-140];DT(5,0,4)rec[-30]_rec[-37]_rec[-40]_rec[-126]_rec[-133]_rec[-136];DT(5,6,4)rec[-29]_rec[-38]_rec[-39]_rec[-125]_rec[-134]_rec[-135];DT(7,3,4)rec[-24]_rec[-33]_rec[-34]_rec[-120]_rec[-129]_rec[-130];DT(7,9,4)rec[-23]_rec[-31]_rec[-32]_rec[-119]_rec[-127]_rec[-128];DT(9,0,4)rec[-18]_rec[-25]_rec[-28]_rec[-114]_rec[-121]_rec[-124];DT(9,6,4)rec[-17]_rec[-26]_rec[-27]_rec[-113]_rec[-122]_rec[-123];DT(11,3,4)rec[-12]_rec[-21]_rec[-22]_rec[-108]_rec[-117]_rec[-118];DT(11,9,4)rec[-11]_rec[-19]_rec[-20]_rec[-107]_rec[-115]_rec[-116];DT(13,4,4)rec[-9]_rec[-10]_rec[-48]_rec[-105]_rec[-106]_rec[-144];DT(13,10,4)rec[-7]_rec[-8]_rec[-47]_rec[-103]_rec[-104]_rec[-143];DT(13,0,4)rec[-6]_rec[-13]_rec[-16]_rec[-102]_rec[-109]_rec[-112];DT(13,6,4)rec[-5]_rec[-14]_rec[-15]_rec[-101]_rec[-110]_rec[-111];DT(15,7,4)rec[-2]_rec[-3]_rec[-41]_rec[-98]_rec[-99]_rec[-137];DT(15,11,4)rec[-1]_rec[-4]_rec[-42]_rec[-97]_rec[-100]_rec[-138];TICK;R_87_93_2_4_8_10_0_6_13_17_19_23_15_21_26_28_32_34_24_30_37_41_43_47_39_45_50_52_56_58_48_54_61_65_67_71_63_69_74_76_80_82_72_78_85_89_91_95;TICK;CY_5_89_11_95_1_0_3_4_7_6_9_10_14_2_20_8_12_13_16_15_18_19_22_21_29_17_35_23_25_24_27_28_31_30_33_34_38_26_44_32_36_37_40_39_42_43_46_45_53_41_59_47_49_48_51_52_55_54_57_58_62_50_68_56_60_61_64_63_66_67_70_69_77_65_83_71_73_72_75_76_79_78_81_82_86_74_92_80_84_85_88_87_90_91_94_93;TICK;CX_3_87_9_93_1_2_5_4_7_8_11_10_12_0_18_6_14_13_16_17_20_19_22_23_27_15_33_21_25_26_29_28_31_32_35_34_36_24_42_30_38_37_40_41_44_43_46_47_51_39_57_45_49_50_53_52_55_56_59_58_60_48_66_54_62_61_64_65_68_67_70_71_75_63_81_69_73_74_77_76_79_80_83_82_84_72_90_78_86_85_88_89_92_91_94_95;TICK;MX_3_9_1_5_7_11_12_18_14_16_20_22_27_33_25_29_31_35_36_42_38_40_44_46_51_57_49_53_55_59_60_66_62_64_68_70_75_81_73_77_79_83_84_90_86_88_92_94;DT(3,2,5)rec[-40]_rec[-42]_rec[-46]_rec[-136]_rec[-138]_rec[-142];DT(3,8,5)rec[-38]_rec[-41]_rec[-44]_rec[-134]_rec[-137]_rec[-140];DT(5,5,5)rec[-33]_rec[-36]_rec[-39]_rec[-129]_rec[-132]_rec[-135];DT(5,11,5)rec[-31]_rec[-35]_rec[-37]_rec[-127]_rec[-131]_rec[-133];DT(7,2,5)rec[-28]_rec[-30]_rec[-34]_rec[-124]_rec[-126]_rec[-130];DT(7,8,5)rec[-26]_rec[-29]_rec[-32]_rec[-122]_rec[-125]_rec[-128];DT(9,5,5)rec[-21]_rec[-24]_rec[-27]_rec[-117]_rec[-120]_rec[-123];DT(9,11,5)rec[-19]_rec[-23]_rec[-25]_rec[-115]_rec[-119]_rec[-121];DT(11,2,5)rec[-16]_rec[-18]_rec[-22]_rec[-112]_rec[-114]_rec[-118];DT(11,8,5)rec[-14]_rec[-17]_rec[-20]_rec[-110]_rec[-113]_rec[-116];DT(13,5,5)rec[-9]_rec[-12]_rec[-15]_rec[-105]_rec[-108]_rec[-111];DT(13,11,5)rec[-7]_rec[-11]_rec[-13]_rec[-103]_rec[-107]_rec[-109];DT(15,2,5)rec[-4]_rec[-6]_rec[-10]_rec[-100]_rec[-102]_rec[-106];DT(15,4,5)rec[-3]_rec[-45]_rec[-48]_rec[-99]_rec[-141]_rec[-144];DT(15,8,5)rec[-2]_rec[-5]_rec[-8]_rec[-98]_rec[-101]_rec[-104];DT(15,10,5)rec[-1]_rec[-43]_rec[-47]_rec[-97]_rec[-139]_rec[-143];TICK;RX_3_9_1_5_7_11_12_18_14_16_20_22_27_33_25_29_31_35_36_42_38_40_44_46_51_57_49_53_55_59_60_66_62_64_68_70_75_81_73_77_79_83_84_90_86_88_92_94;TICK;CX_3_87_9_93_1_2_5_4_7_8_11_10_12_0_18_6_14_13_16_17_20_19_22_23_27_15_33_21_25_26_29_28_31_32_35_34_36_24_42_30_38_37_40_41_44_43_46_47_51_39_57_45_49_50_53_52_55_56_59_58_60_48_66_54_62_61_64_65_68_67_70_71_75_63_81_69_73_74_77_76_79_80_83_82_84_72_90_78_86_85_88_89_92_91_94_95;TICK;XCY_85_1_91_7_2_3_6_5_8_9_0_11_4_16_10_22_15_14_17_18_21_20_23_12_13_25_19_31_26_27_30_29_32_33_24_35_28_40_34_46_39_38_41_42_45_44_47_36_37_49_43_55_50_51_54_53_56_57_48_59_52_64_58_70_63_62_65_66_69_68_71_60_61_73_67_79_74_75_78_77_80_81_72_83_76_88_82_94_87_86_89_90_93_92_95_84;TICK;MY_87_93_2_4_8_10_0_6_13_17_19_23_15_21_26_28_32_34_24_30_37_41_43_47_39_45_50_52_56_58_48_54_61_65_67_71_63_69_74_76_80_82_72_78_85_89_91_95;DT(3,5,6)rec[-39]_rec[-41]_rec[-45]_rec[-135]_rec[-137]_rec[-141];DT(3,11,6)rec[-37]_rec[-42]_rec[-43]_rec[-133]_rec[-138]_rec[-139];DT(5,2,6)rec[-34]_rec[-36]_rec[-40]_rec[-130]_rec[-132]_rec[-136];DT(5,8,6)rec[-32]_rec[-35]_rec[-38]_rec[-128]_rec[-131]_rec[-134];DT(7,5,6)rec[-27]_rec[-29]_rec[-33]_rec[-123]_rec[-125]_rec[-129];DT(7,11,6)rec[-25]_rec[-30]_rec[-31]_rec[-121]_rec[-126]_rec[-127];DT(9,2,6)rec[-22]_rec[-24]_rec[-28]_rec[-118]_rec[-120]_rec[-124];DT(9,8,6)rec[-20]_rec[-23]_rec[-26]_rec[-116]_rec[-119]_rec[-122];DT(11,5,6)rec[-15]_rec[-17]_rec[-21]_rec[-111]_rec[-113]_rec[-117];DT(11,11,6)rec[-13]_rec[-18]_rec[-19]_rec[-109]_rec[-114]_rec[-115];DT(13,2,6)rec[-10]_rec[-12]_rec[-16]_rec[-106]_rec[-108]_rec[-112];DT(13,8,6)rec[-8]_rec[-11]_rec[-14]_rec[-104]_rec[-107]_rec[-110];DT(15,1,6)rec[-4]_rec[-46]_rec[-48]_rec[-100]_rec[-142]_rec[-144];DT(15,5,6)rec[-3]_rec[-5]_rec[-9]_rec[-99]_rec[-101]_rec[-105];DT(15,7,6)rec[-2]_rec[-44]_rec[-47]_rec[-98]_rec[-140]_rec[-143];DT(15,11,6)rec[-1]_rec[-6]_rec[-7]_rec[-97]_rec[-102]_rec[-103];TICK;RY_87_93_2_4_8_10_0_6_13_17_19_23_15_21_26_28_32_34_24_30_37_41_43_47_39_45_50_52_56_58_48_54_61_65_67_71_63_69_74_76_80_82_72_78_85_89_91_95;TICK;XCY_85_1_91_7_2_3_6_5_8_9_0_11_4_16_10_22_15_14_17_18_21_20_23_12_13_25_19_31_26_27_30_29_32_33_24_35_28_40_34_46_39_38_41_42_45_44_47_36_37_49_43_55_50_51_54_53_56_57_48_59_52_64_58_70_63_62_65_66_69_68_71_60_61_73_67_79_74_75_78_77_80_81_72_83_76_88_82_94_87_86_89_90_93_92_95_84;TICK;CY_89_5_95_11_0_1_4_3_6_7_10_9_2_14_8_20_13_12_15_16_19_18_21_22_17_29_23_35_24_25_28_27_30_31_34_33_26_38_32_44_37_36_39_40_43_42_45_46_41_53_47_59_48_49_52_51_54_55_58_57_50_62_56_68_61_60_63_64_67_66_69_70_65_77_71_83_72_73_76_75_78_79_82_81_74_86_80_92_85_84_87_88_91_90_93_94;TICK;M_3_9_1_5_7_11_12_18_14_16_20_22_27_33_25_29_31_35_36_42_38_40_44_46_51_57_49_53_55_59_60_66_62_64_68_70_75_81_73_77_79_83_84_90_86_88_92_94;DT(3,4,7)rec[-39]_rec[-40]_rec[-48]_rec[-135]_rec[-136]_rec[-144];DT(3,10,7)rec[-37]_rec[-38]_rec[-47]_rec[-133]_rec[-134]_rec[-143];DT(5,7,7)rec[-32]_rec[-33]_rec[-41]_rec[-128]_rec[-129]_rec[-137];DT(5,11,7)rec[-31]_rec[-34]_rec[-42]_rec[-127]_rec[-130]_rec[-138];DT(7,4,7)rec[-27]_rec[-28]_rec[-36]_rec[-123]_rec[-124]_rec[-132];DT(7,10,7)rec[-25]_rec[-26]_rec[-35]_rec[-121]_rec[-122]_rec[-131];DT(9,7,7)rec[-20]_rec[-21]_rec[-29]_rec[-116]_rec[-117]_rec[-125];DT(9,11,7)rec[-19]_rec[-22]_rec[-30]_rec[-115]_rec[-118]_rec[-126];DT(11,4,7)rec[-15]_rec[-16]_rec[-24]_rec[-111]_rec[-112]_rec[-120];DT(11,10,7)rec[-13]_rec[-14]_rec[-23]_rec[-109]_rec[-110]_rec[-119];DT(13,7,7)rec[-8]_rec[-9]_rec[-17]_rec[-104]_rec[-105]_rec[-113];DT(13,11,7)rec[-7]_rec[-10]_rec[-18]_rec[-103]_rec[-106]_rec[-114];DT(15,0,7)rec[-6]_rec[-43]_rec[-46]_rec[-102]_rec[-139]_rec[-142];DT(15,6,7)rec[-5]_rec[-44]_rec[-45]_rec[-101]_rec[-140]_rec[-141];DT(15,4,7)rec[-3]_rec[-4]_rec[-12]_rec[-99]_rec[-100]_rec[-108];DT(15,10,7)rec[-1]_rec[-2]_rec[-11]_rec[-97]_rec[-98]_rec[-107];TICK;R_3_9_1_5_7_11_12_18_14_16_20_22_27_33_25_29_31_35_36_42_38_40_44_46_51_57_49_53_55_59_60_66_62_64_68_70_75_81_73_77_79_83_84_90_86_88_92_94;TICK;CY_89_5_95_11_0_1_4_3_6_7_10_9_2_14_8_20_13_12_15_16_19_18_21_22_17_29_23_35_24_25_28_27_30_31_34_33_26_38_32_44_37_36_39_40_43_42_45_46_41_53_47_59_48_49_52_51_54_55_58_57_50_62_56_68_61_60_63_64_67_66_69_70_65_77_71_83_72_73_76_75_78_79_82_81_74_86_80_92_85_84_87_88_91_90_93_94;TICK;H_0_1_2_3_4_5_6_7_8_9_10_11_12_13_14_15_16_17_18_19_20_21_22_23_24_25_26_27_28_29_30_31_32_33_34_35_36_37_38_39_40_41_42_43_44_45_46_47_48_49_50_51_52_53_54_55_56_57_58_59_60_61_62_63_64_65_66_67_68_69_70_71_72_73_74_75_76_77_78_79_80_81_82_83_84_85_86_87_88_89_90_91_92_93_94_95;TICK;M_0_1_2_3_4_5_6_7_8_9_10_11_12_13_14_15_16_17_18_19_20_21_22_23_24_25_26_27_28_29_30_31_32_33_34_35_36_37_38_39_40_41_42_43_44_45_46_47_48_49_50_51_52_53_54_55_56_57_58_59_60_61_62_63_64_65_66_67_68_69_70_71_72_73_74_75_76_77_78_79_80_81_82_83_84_85_86_87_88_89_90_91_92_93_94_95;DT(3,1,8)rec[-82]_rec[-83]_rec[-84]_rec[-94]_rec[-95]_rec[-96]_rec[-184]_rec[-186]_rec[-190];DT(3,4,8)rec[-80]_rec[-81]_rec[-82]_rec[-92]_rec[-93]_rec[-94];DT(3,7,8)rec[-76]_rec[-77]_rec[-78]_rec[-88]_rec[-89]_rec[-90]_rec[-182]_rec[-185]_rec[-188];DT(3,10,8)rec[-74]_rec[-75]_rec[-76]_rec[-86]_rec[-87]_rec[-88];DT(5,4,8)rec[-67]_rec[-68]_rec[-69]_rec[-79]_rec[-80]_rec[-81]_rec[-177]_rec[-180]_rec[-183];DT(5,7,8)rec[-65]_rec[-66]_rec[-67]_rec[-77]_rec[-78]_rec[-79];DT(5,11,8)rec[-61]_rec[-71]_rec[-72]_rec[-73]_rec[-83]_rec[-84];DT(5,10,8)rec[-61]_rec[-62]_rec[-63]_rec[-73]_rec[-74]_rec[-75]_rec[-175]_rec[-179]_rec[-181];DT(7,1,8)rec[-58]_rec[-59]_rec[-60]_rec[-70]_rec[-71]_rec[-72]_rec[-172]_rec[-174]_rec[-178];DT(7,4,8)rec[-56]_rec[-57]_rec[-58]_rec[-68]_rec[-69]_rec[-70];DT(7,7,8)rec[-52]_rec[-53]_rec[-54]_rec[-64]_rec[-65]_rec[-66]_rec[-170]_rec[-173]_rec[-176];DT(7,10,8)rec[-50]_rec[-51]_rec[-52]_rec[-62]_rec[-63]_rec[-64];DT(9,4,8)rec[-43]_rec[-44]_rec[-45]_rec[-55]_rec[-56]_rec[-57]_rec[-165]_rec[-168]_rec[-171];DT(9,7,8)rec[-41]_rec[-42]_rec[-43]_rec[-53]_rec[-54]_rec[-55];DT(9,11,8)rec[-37]_rec[-47]_rec[-48]_rec[-49]_rec[-59]_rec[-60];DT(9,10,8)rec[-37]_rec[-38]_rec[-39]_rec[-49]_rec[-50]_rec[-51]_rec[-163]_rec[-167]_rec[-169];DT(11,1,8)rec[-34]_rec[-35]_rec[-36]_rec[-46]_rec[-47]_rec[-48]_rec[-160]_rec[-162]_rec[-166];DT(11,4,8)rec[-32]_rec[-33]_rec[-34]_rec[-44]_rec[-45]_rec[-46];DT(11,7,8)rec[-28]_rec[-29]_rec[-30]_rec[-40]_rec[-41]_rec[-42]_rec[-158]_rec[-161]_rec[-164];DT(11,10,8)rec[-26]_rec[-27]_rec[-28]_rec[-38]_rec[-39]_rec[-40];DT(13,4,8)rec[-19]_rec[-20]_rec[-21]_rec[-31]_rec[-32]_rec[-33]_rec[-153]_rec[-156]_rec[-159];DT(13,7,8)rec[-17]_rec[-18]_rec[-19]_rec[-29]_rec[-30]_rec[-31];DT(13,11,8)rec[-13]_rec[-23]_rec[-24]_rec[-25]_rec[-35]_rec[-36];DT(13,10,8)rec[-13]_rec[-14]_rec[-15]_rec[-25]_rec[-26]_rec[-27]_rec[-151]_rec[-155]_rec[-157];DT(15,1,8)rec[-10]_rec[-11]_rec[-12]_rec[-22]_rec[-23]_rec[-24]_rec[-148]_rec[-150]_rec[-154];DT(15,4,8)rec[-8]_rec[-9]_rec[-10]_rec[-20]_rec[-21]_rec[-22];DT(15,5,8)rec[-7]_rec[-8]_rec[-9]_rec[-91]_rec[-92]_rec[-93]_rec[-147]_rec[-189]_rec[-192];DT(15,7,8)rec[-5]_rec[-6]_rec[-7]_rec[-89]_rec[-90]_rec[-91];DT(13,6,8)rec[-4]_rec[-5]_rec[-6]_rec[-16]_rec[-17]_rec[-18]_rec[-146]_rec[-149]_rec[-152];DT(15,10,8)rec[-2]_rec[-3]_rec[-4]_rec[-14]_rec[-15]_rec[-16];DT(15,11,8)rec[-1]_rec[-11]_rec[-12]_rec[-85]_rec[-95]_rec[-96];DT(1,10,8)rec[-1]_rec[-2]_rec[-3]_rec[-85]_rec[-86]_rec[-87]_rec[-145]_rec[-187]_rec[-191];OI(0)rec[-11]_rec[-12]_rec[-23]_rec[-24]_rec[-35]_rec[-36]_rec[-47]_rec[-48]_rec[-59]_rec[-60]_rec[-71]_rec[-72]_rec[-83]_rec[-84]_rec[-95]_rec[-96]_rec[-150]_rec[-162]_rec[-174]_rec[-186]_rec[-244]_rec[-256]_rec[-268]_rec[-280]_rec[-340]_rec[-342]_rec[-352]_rec[-354]_rec[-364]_rec[-366]_rec[-376]_rec[-378]_rec[-438]_rec[-450]_rec[-462]_rec[-474]_rec[-532]_rec[-544]_rec[-556]_rec[-568]_rec[-628]_rec[-630]_rec[-640]_rec[-642]_rec[-652]_rec[-654]_rec[-664]_rec[-666]}{in Crumble}).

\edit{We have numerically verified this behavior using STIM~\cite{gidney2021stim}. For fixed $d$, we generate circuits with a varying number $N_m$ of measurement rounds and produce their detector error models. We can then numerically find the minimum distance between detectors occuring around the beginning of the circuit (some fixed distance $\delta N_m$ from $0$) and detectors occuring around the end of the circuit (some fixed distance $\delta N_m$ from $N_m$). The precise value of $\delta N_m$ is not particularly important, as long as we keep it fixed as we scale $N_m$. We can then consider two quantities: (1) $d_\text{graph}$, the minimum number of graphlike errors it takes to connect the beginning detectors to end detectors (2) $d_\text{hyper}$, the minimum number of graphlike errors connecting the detectors if we decompose all hyperedges $h$ into complete graphs on the degree-$|h|$ vertices of $h$. These distances provide upper/lower bounds on the timelike distance $d_t$: $d_\text{graph}\geq d_t\geq d_\text{hyper}$. For the standard circuit we find $d_\text{graph}=d_\text{hyper}=3N_m/4$, so that we require $N_m=4d_t/3$ to reach distance $d_t$. For the dynamic circuit, we find $d_\text{graph}=2N_m, d_\text{hyper}=N_m$, so we require at most $N_m=d_t$ to reach distance $d_t$ (but $N_m=d_t/2$ may be sufficient). Establishing the precise relationship between $d_t$ and $N_m$ for the dynamic circuit will require a stability simulation~\cite{gidney2022stability}.
}

\section{Additional numerical results}
\label{appendix:numerics}
Here, we include a plot of our extrapolation of logical error rate as a function of distance, used to compute the teraquop footprint for $p=10^{-3}$ \edit{as well as} graphs of the separate horizontal and vertical logical error rates, rather than summed as in the main text. \edit{We present results from both MWPM decoding, as well as the more accurate but slower correlated matching decoding~\cite{fowler2013optimal}.} The V logical error probability is the probability that the vertical logical operator is incorrect, i.e. the probability that an undetected horizontal logical operator has been applied, and similarly for the H logical error probability. The larger V logical error probability for the standard circuit was also seen in~\cite{gidney2022benchmarking}. \edit{The steeper slope of the V logical error probability in the dynamic circuit results from a complicated issue of error multiplicity that is difficult to fully account for. We note that all of these probabilities have a slope steeper than $p^{\lfloor d/2\rfloor}$ despite the circuits having a numerically-verified graphlike distance of $d$, due to higher-distance errors dominating at these values of $p$.}

\begin{figure*}
    \includegraphics[width=\columnwidth]{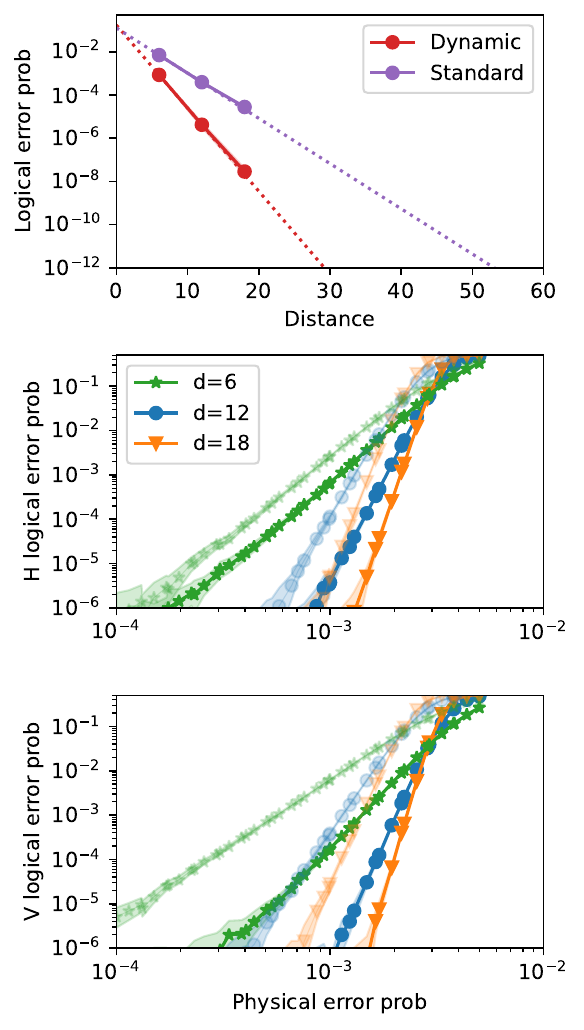}
    \includegraphics[width=\columnwidth]{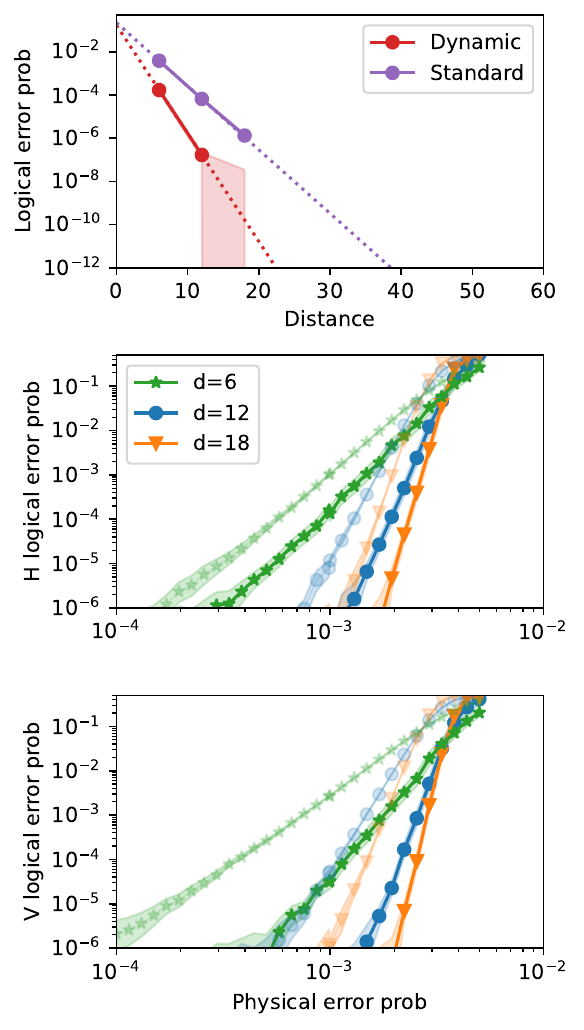}
    \caption{Our extrapolation to the teraquop regime for $p=10^{-3}$, and the logical error rates separated into horizontal and vertical error rates, \edit{for (left) MWPM and (right) correlated matching. Circuits simulated with STIM~\cite{gidney2021stim} and decoded with PyMatching~\cite{higgott2025sparse}}.}
    \label{fig:extended_data}
\end{figure*}

\bibliography{main.bib}

\end{document}